\documentclass[aps,prl,twocolumn,showpacs]{revtex4}
\usepackage{graphicx}
\usepackage{bm}
\begin{document}
\title{Hot electrons in low-dimensional phonon systems}
\author{S.-X. Qu,$^1$  A. N. Cleland,$^2$ and M. R. Geller$^1$}
\affiliation{$^1$Department of Physics and Astronomy, University of Georgia, Athens, Georgia 30602-2451}
\affiliation{$^2$Department of Physics, University of California, Santa Barbara, California 93106}

\date{March 15, 2005}

\begin{abstract}
A simple bulk model of electron-phonon coupling in metals has been surprisingly successful in explaining experiments on metal films that actually involve surface- or other low-dimensional phonons. However, by an exact application of this standard model to a semi-infinite substrate with a free surface, making use of the actual vibrational modes of the substrate, we show that such agreement is fortuitous, and that the model actually predicts a low-temperature crossover from the familiar $T^5$ temperature dependence to a stronger $T^6 \log T$ scaling. Comparison with existing experiments suggests a widespread breakdown of the standard model of electron-phonon thermalization in metals.
\end{abstract}

\maketitle

The coupling between electrons and phonons plays a crucial role in determining the thermal properties of nanostructures. The widely used ``standard" model of low temperature electron-phonon thermal coupling and hot-electron effects in bulk metals \cite{GantmakherRPP74,WellstoodPRB94} assumes (i) a clean three-dimensional free-electron gas with a spherical Fermi surface, rapidly equilibrated to a temperature $T_{\rm el}$; (ii) a continuum description of the acoustic phonons, which have a temperature $T_{\rm ph}$; (iii) a negligible Kapitza-like thermal boundary resistance \cite{Little59} between the metal and any surrounding dielectric, an assumption that is often well justified experimentally; and (iv), a deformation-potential electron-phonon coupling, expected to be the dominant interaction at long-wavelengths. In a bulk metal, the net rate $P$ of thermal energy transfer between the electron and phonon subsystems is \cite{WellstoodPRB94}
\begin{equation}
P = \Sigma V_{\rm el} \big(T_{\rm el}^5 - T_{\rm ph}^5 \big),
\label{Wellstood formula}
\end{equation}
where $V_{\rm el}$ is the volume of the metal, and
\begin{equation}
\Sigma \equiv { 8\, \zeta(5) \, k_{\rm B}^5 \,  \epsilon_{\rm F}^2 \, N_{\rm el}(\epsilon_{\rm F}) \,  \over 3 \pi  \hbar^4  \rho v_{\rm F} v_{\rm l}^4}.
\label{Wellstood coefficient}
\end{equation}
Here $\zeta$ is the Riemann zeta function, $\epsilon_{\rm F}$ is the Fermi energy, $N_{\rm el}$ is the electronic density of states (DOS) per unit volume, $\rho$ is the mass density, $v_{\rm l}$ is the bulk longitudinal sound speed, and $v_{\rm F}$ is the Fermi velocity.

\begin{figure}
\includegraphics[width=10.0cm]{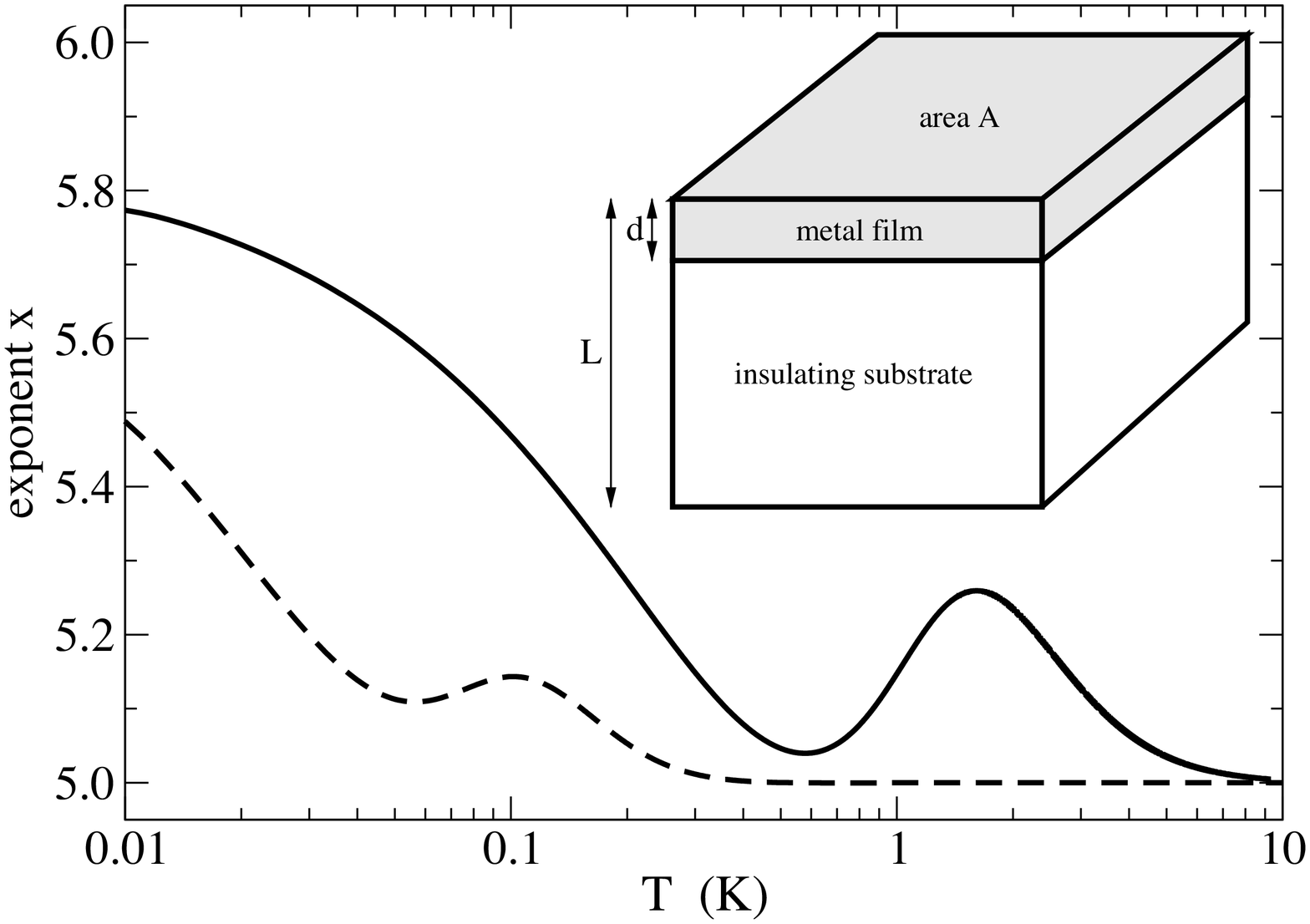}
\caption{(inset) Conducting film of thickness $d$ attached to insulator. The top surface of the metal is stress-free. (main) Temperature dependence of the thermal power exponent $x$ for a $10 \, {\rm nm}$ (solid curve) and $100 \, {\rm nm}$ (dashed curve) Cu film.}
\label{exponent figure}
\end{figure}

This model, which has no adjustable parameters, has successfully explained some experiments \cite{WellstoodPRB94,RoukesPRL85,YungAPL02}, but others report a power-law temperature dependence with smaller exponents \cite{LiuPRB91,DiTusaPRL92}, indicating an enhanced electron-phonon coupling at low temperatures. However, the experiments typically involve heating measurements in thin metal films deposited on semiconducting or insulating substrates, and the relevant phonons at low temperature are strongly modified by the exposed stress-free surface. An attempt to directly probe such phonon-dimensionality effects was carried out by DiTusa {\it et al.} \cite{DiTusaPRL92}, who intentionally suspended some of their samples, necessarily modifying the vibrational spectrum, although they found no significant difference from their supported films. We argue that the paradox reported in Ref.~\cite{DiTusaPRL92} is actually quite widespread, and all experiments known to us on supported films actually contradict the standard model when that model is modified to account for the actual vibrational modes present in a realistic supported-film geometry, illustrated in Fig.~\ref{exponent figure}. Our results have important implications for the thermal properties of mesoscopic and low-dimensional phonon systems and the use of such systems as nanoscale thermometers, bolometers, and calorimeters \cite{Roukes99,SchwabNat00,SchmidtPRL04}.

The Hamiltonian we consider (suppressing spin) is $H = \sum_{\bf k} \epsilon_{\bf k} \, c_{\bf k}^\dagger c_{\bf k} + \sum_n \hbar \omega_n \, a_n^\dagger a_n + \delta H,$ where $c_{\bf k}^\dagger$ and $c_{\bf k}$ are electron creation and annihilation operators, with ${\bf k}$ the momentum, and $a_n^\dagger$ and $a_n$ are bosonic phonon creation and annihilation operators. The vibrational modes, labeled by an index $n$, are eigenfunctions of the continuum elasticity equation $v_{\rm t}^2 {\bm \nabla} \times {\bm \nabla} \times {\bf u} - v_{\rm l}^2 {\bm \nabla} ({\bm \nabla} \cdot {\bf u}) = \omega^2  {\bf u} $ for linear isotropic media, along with accompanying boundary conditions. $v_{\rm t}$ and $v_{\rm l}$ are the bulk transverse and longitudinal sound velocities. $\delta H \equiv {\textstyle{2 \over 3}} \epsilon_{\rm F} \! \int_{V_{\rm el}} \! \!  d^3r \, \psi^\dagger \psi \, {\bf \nabla} \cdot {\bf u}$ is the deformation-potential electron-phonon interaction, with ${\bf u}({\bf r}) = \sum_n (2 \rho \omega_n)^{-{1 \over 2}} [ {\bf f}_n({\bf r}) \, a_n +  {\bf f}^*_n({\bf r}) \, a_n^\dagger ]$ the quantized displacement field. The vibrational eigenfunctions ${\bf f}_n({\bf r})$ are defined to be solutions of the elasticity field equations, normalized over the phonon volume $V_{\rm ph}$ according to $\int_{V_{\rm ph}} \! \!  d^3r \ {\bf f}_n^* \cdot {\bf f}_{n'} = \delta_{nn'}.$ It will be convenient to rewrite the electron-phonon interaction as $\delta H = \sum_{{\bf k q}n} [g_{n {\bf q}} \, c_{\bf k+q}^\dagger c_{\bf k} \, a_n + g_{n {\bf q}}^* \, c_{\bf k-q}^\dagger c_{\bf k} \, a_n^\dagger ],$ with coupling constant $g_{n {\bf q}} \equiv \frac{2}{3} \epsilon_{\rm F} (2 \rho \omega_n)^{-{1\over2}} V_{\rm el} ^{-1} \! \int_{V_{\rm el}} \! \!  d^3r \ {\bf \nabla} \cdot {\bf f}_n \, e^{-i{\bf q} \cdot {\bf r}}.$ Note that we allow for different electron and phonon volumes.

The quantity we calculate is the thermal energy per unit time transferred from the electrons to the phonons,
\begin{equation}
P \equiv 2 \sum_{{\bf k q}n} \hbar \omega_n \big[ \Gamma^{\rm em}_n({\bf k} \rightarrow {\bf k} - {\bf q}) - \Gamma^{\rm ab}_n({\bf k} \rightarrow {\bf k} + {\bf q}) \big],
\label{P definition}
\end{equation} 
where
\begin{eqnarray}
&& \Gamma^{\rm em}_n({\bf k} \rightarrow {\bf k} - {\bf q}) = 2 \pi \, |g_{n {\bf q}}|^2 \, [n_{\rm B}(\omega_n) + 1]  \nonumber \\
&\times&  n_{\rm F}(\epsilon_{\bf k}) [1-n_{\rm F}(\epsilon_{{\bf k}-{\bf q}}) ] \, \delta( \epsilon_{{\bf k}-{\bf q}} - \epsilon_{\bf k} +  \omega_n ) 
\end{eqnarray}
is the golden-rule rate for an electron of momentum ${\bf k}$ to scatter to ${\bf k} - {\bf q}$ while emitting a phonon $n$, and  
\begin{eqnarray}
&&\Gamma^{\rm ab}_n({\bf k} \rightarrow {\bf k} + {\bf q}) = 2 \pi \, |g_{n {\bf q}}|^2 \, n_{\rm B}(\omega_n) \nonumber \\
&\times& n_{\rm F}(\epsilon_{\bf k}) \, [1-n_{\rm F}(\epsilon_{{\bf k}+{\bf q}}) ] \, \delta( \epsilon_{{\bf k}+{\bf q}} - \epsilon_{\bf k} - \omega_n )
\end{eqnarray}
is the corresponding phonon absorption rate. $n_{\rm B}$ is the Bose distribution function with temperature $T_{\rm ph}$ and $n_{\rm F}$ is the Fermi distribution with temperature $T_{\rm el}$. The factor of 2 in (\ref{P definition}) accounts for spin degeneracy. It is possible to obtain an exact expression for $P;$ the result (suppressing factors of $\hbar$ and $k_{\rm B}$) is
\begin{eqnarray*}
&& {\hskip -0.20in} P = {m^2 V_{\rm el}^2 \over 8 \pi^4} \sum_{n}  \int_0^{\infty} \! \! \! d\omega \, \delta(\omega - \omega_n) \, \big( {\textstyle{\omega \over e^{\omega/T_{\rm el}}-1}} - {\textstyle{\omega \over e^{\omega/ T_{\rm ph}}-1}} \big) \nonumber \\
&\times& {\hskip -0.10in} \int d^3k \, {|g_{n {\bf k}}|^2 \over |{\bf k}|} \bigg[ \omega + T_{\rm el} \ln \bigg( {\textstyle{ 1 +  \exp[({m \omega^2 \over 2 k^2} 
+ {k^2 \over 8m} - {\omega \over 2} - \mu) / T_{\rm el}])  \over  1 +  \exp[({m \omega^2 \over 2 k^2} + {k^2 \over 8m} + {\omega \over 2} - \mu) / T_{\rm el}]) }} 
\bigg) \bigg]. {\hskip 0.20in}
\label{general thermal power}
\end{eqnarray*}

The logarithmic term in $P$ can be shown to be negligible in the temperature regime of interest and will be dropped. Carrying out the ${\bf k}$ integration then leads to
\begin{equation}
P \! = \! {v_{\rm l}^4 \, \Sigma \, V_{\rm el} \over24 \, \zeta(5)} \! \! \int_0^{\omega_{\rm D}} \! \! \! \! \! d\omega \, F(\omega) \bigg({\omega  \over e^{\omega/T_{\rm el}}-1} - 
{\omega  \over e^{\omega/T_{\rm ph}}-1} \bigg),
\label{thermal power formula}
\end{equation}
where 
$F(\omega) \equiv \sum_{n} U_n \, \delta(\omega - \omega_n)$ is a strain-weighted vibrational DOS, with
\begin{equation}
U_n \equiv { 1 \over V_{\rm el}} \int_{V_{\rm el}} \! d^3r \, d^3r' \ {{\bm \nabla} \cdot {\bf f}_n({\bf r}) \ {\bm \nabla}' \cdot {\bf f}_n^*({\bf r}') \over |{\bf r}-{\bf r}'|^2 + a^2}.
\label{effective interaction}
\end{equation}
Here $\omega_{\rm D}$ is the Debye frequency. $U_n$ can be interpreted as an energy associated with mass-density fluctuations interacting via an inverse-square potential \cite{densitynote}, cut off at distances of the order of the lattice constant $a$. We have reduced the calculation of $P$ to the calculation of $F(\omega)$. Allen \cite{AllenPRL87} has derived a related weighted-DOS formalism.

We now calculate $F(\omega)$ and $P$ for a metal film of thickness $d$ attached to the free surface of an isotropic elastic continuum with $L \rightarrow \infty;$ see the inset to Fig.~\ref{exponent figure}. For calculational simplicity, the film and substrate are assumed to have the same elastic parameters, characterized by a mass density $\rho$ and bulk sound velocities $v_{\rm t}$ and $v_{\rm l}$. Where material parameters are necessary we shall assume a Cu film; however, the qualitative behavior we obtain is generic. The evaluation of $F(\omega)$ requires the vibrational eigenfunctions for a semi-infinite substrate with a free surface, which have been obtained in the classic paper by Ezawa \cite{Ezawa71}. The modes are labeled by a branch index $m$, taking the five values \ SH, $+$, $-$, 0, and R, by a two-dimensional wavevector ${\bf K}$ in the plane defined by the surface, and by a parameter $c$ with the dimensions of velocity that is continuous for all branches except the Rayleigh branch $m \! = \! {\rm R}$. With the normalization convention of Ref.~\cite{Ezawa71} we have
\begin{eqnarray}
F(\omega) &=& \sum_{\bf K} U_{{\rm R} {\bf K}} \ \delta(\omega - c_{\rm R} K) \nonumber \\
&+&  \sum_{m \neq {\rm R}} \sum_{\bf K} \int \! dc \ U_{m {\bf K} c} \ \delta(\omega - c K).
\label{continuous weighted DOS}
\end{eqnarray}
The range of the parameter $c$ depends on the branch $m$, and is summarized in Table \ref{c table}. The frequency of mode $ m {\bf K} c$ is $c K.$ 

\begin{table}[!h]
{\centering
\begin{tabular}{|c|c|}  \hline
 $m$  & \ \ \ \ \ \ range of $c$ \ \ \ \ \   \\ \hline
SH    & $[v_{\rm t} , \infty]$  \\
$\pm$ & $[v_{\rm l} , \infty]$  \\
0     & $[v_{\rm t} , v_{\rm l}]$  \\
R     & \, $c_{\rm R}$ \, (discrete)  \\  \hline
\end{tabular}
\par}
\caption{Values of the parameter $c$ for the five branches of vibrational modes of a semi-infinite substrate.}
\label{c table}
\end{table}

Turning to an evaluation of (\ref{continuous weighted DOS}), the SH branch is purely transverse, so $U_{\rm SH} = 0.$ The normalized eigenmodes for the $\pm$ branches are 
\begin{eqnarray}
{\bf f}_{\pm} &=& \sqrt{K \over 4 \pi c A} \bigg\{ \bigg[ \mp \alpha^{-{1 \over 2}} \, \big( e^{-i \alpha K z} - \zeta_{\pm} \, e^{i \alpha K z} \big) + i \beta^{1 \over 2}  \nonumber \\ 
&\times&  \big( e^{-i \beta K z} + \zeta_{\pm} \, e^{i \beta K z} \big) \bigg] {\bf e}_{\rm K} 
+ \bigg[ \pm \alpha^{1 \over 2} \big( e^{-i \alpha K z} + \zeta_{\pm} \, e^{i \alpha K z} \big) \nonumber \\
&+& i \beta^{-{1 \over 2}} \big( e^{-i \beta K z} - \zeta_{\pm} \, e^{i \beta K z} \big) \bigg] {\bf e}_z \bigg\} e^{i {\bf K} \cdot {\bf r}},   
\label{pm modes}
\end{eqnarray}
where $\alpha \! \equiv \! \sqrt{(c/v_{\rm l})^2 - 1}$ and $\beta \! \equiv \! \sqrt{(c/v_{\rm t})^2 - 1}.$ Here 
\begin{equation}
\zeta_{\pm} \equiv {[(\beta^2 - 1) \pm 2 i \sqrt{\alpha \beta} ]^2 \over (\beta^2 -1)^2 + 4 \alpha \beta}, \ \ \ \ {\rm with} \ \ \ \ |\zeta_{\pm}|=1.
\end{equation}
Then
\begin{equation}
{\bm \nabla} \cdot {\bf f}_{\pm} = \mp i {c^{3 \over 2} K^{3 \over 2} \over \sqrt{4 \pi \alpha A} \, v_{\rm l}^2 } \,
\big(e^{-i \alpha K z} - \zeta_\pm \, e^{i \alpha K z} \big) \, e^{i {\bf K} \cdot {\bf r}} 
\label{pm divergence}
\end{equation}
and $U_{\pm} = (c^3 K / \alpha \, v_{\rm l}^4 V_{\rm el} ) \, I_{\pm}(Kd,c),$ where
\begin{eqnarray*}
I_{\pm}(Z,c) \! &\equiv& \! {\rm Re} \! \int_0^Z \! \! dx \, dx' \, {\sf K}_0\big( \sqrt{(x \! - \! x')^2 + a^2Z^2/d^2 } \big) \nonumber \\
&\times& \! \big[ e^{i \alpha(x-x')} - \zeta_\pm e^{i \alpha (x+x')} \big]. 
\end{eqnarray*}
${\sf K}_0$ is a modified Bessel function. To obtain $U_{\pm}$ we use translational invariance in the $xy$ plane to write (\ref{effective interaction}) as
\begin{eqnarray}
U_{m {\bf K} c} &=& {A \over V_{\rm el}}  \int_0^d dz \, dz' \int_A d^2R  \nonumber \\ 
&\times& {{\bm \nabla} \cdot {\bf f}_{m {\bf K} c}({\bf R},z) \ {\bm \nabla}' \cdot {\bf f}_{m {\bf K} c}^* (0, z') \over R^2 + (z-z')^2 + a^2},
\label{continuous effective interaction}
\end{eqnarray}
where ${\bf R} \equiv (x,y)$ is a two-dimensional coordinate vector. Then we scale out $K$, do the angular integration, and use the identity $\int_0^\infty dR \, R \, {\sf J}_0(R) \, [R^2 + s^2]^{-1} = {\sf K}_0(|s|),$ where ${\sf J}_0$ is a Bessel function of the first kind. 

Next we consider the $m=0$ branch, for which 
\begin{eqnarray}
{\bf f}_{0} &=& \sqrt{K \over 2 \pi \beta c A} \bigg\{ \bigg[ i {\cal C} \, e^{- \gamma K z} + i \beta \, e^{-i \beta K z} + i \beta {\cal A} \, e^{i \beta K z} \bigg]  {\bf e}_{\rm K} \nonumber \\ &+& \bigg[ - \gamma {\cal C} \, e^{- \gamma K z}  + i e^{-i \beta K z} - i {\cal A} \, e^{i \beta K z} \bigg] {\bf e}_z \bigg\} e^{i {\bf K} \cdot {\bf r}} \! , \ \ 
\label{0 modes}
\end{eqnarray}
where $\gamma \equiv \sqrt{1 - (c/v_{\rm l})^2},$
$$
{\cal A} \equiv {(\beta^2 - 1)^2 - 4 i \beta \gamma \over (\beta^2 - 1)^2 + 4 i \beta \gamma}, \ \ \ {\rm and} \ \ \ {\cal C}\equiv { 4 \beta (\beta^2 - 1) \over (\beta^2 - 1)^2 + 4 i \beta \gamma}.
$$
Then
\begin{equation}
{\bm \nabla} \cdot {\bf f}_{0} = - {c^{3 \over 2} K^{3 \over 2} {\cal C} \over \sqrt{2 \pi \beta A} \, v_{\rm l}^2 } 
\ e^{-\gamma K z} \, e^{i {\bf K} \cdot {\bf r}}
\label{0 divergence}
\end{equation}
and $U_{0} = (|{\cal C}|^2 c^3 K / \beta \, v_{\rm l}^4 V_{\rm el}) \, I_0(Kd,c),$ where
$$
I_0(Z,c) \! \equiv \! \int_0^Z \! \! dx \, dx' \, {\sf K}_0\big( \sqrt{(x \! - \! x')^2 + a^2Z^2/d^2 } \big) e^{-\gamma (x+x')}.
$$

Finally, for the Rayleigh branch,
\begin{eqnarray}
{\bf f}_{\rm R} &=& \sqrt{K \over {\cal K}A} \bigg\{ \bigg[ i e^{-\varphi K z} - i \big({\textstyle{2 \varphi \eta \over 1 + 
\eta^2}}\big) \, e^{- \eta K z} \bigg] {\bf e}_{\rm K} \nonumber \\
&-& \bigg[ \varphi e^{- \varphi K z} - \big({\textstyle{2 \varphi \over 1 + \eta^2}} \big) \, e^{- \eta K z} \bigg] {\bf e}_z \bigg\} e^{i {\bf K} \cdot {\bf r}},
\label{R modes}
\end{eqnarray}
where $\varphi \equiv \sqrt{1 - (c_{\rm R}/v_{\rm l})^2}, \ \eta \equiv \sqrt{1 - (c_{\rm R}/v_{\rm t})^2},$ and ${\cal K} \equiv (\varphi - \eta)(\varphi - \eta + 2 \varphi \eta^2) / 2 \varphi \eta^2.$ $c_{\rm R} $ is the velocity of the Rayleigh surface waves, given by $c_{\rm R}= \xi \, v_{\rm t}$, where $\xi$ is the root between 0 and 1 of
$\xi^6 - 8 \xi^4 + 8(3-2\nu^2) \xi^2 - 16(1-\nu^2),$ with $ \nu \equiv v_{\rm t}/ v_{\rm l}$. For Cu, $\nu = 0.52$ and $\xi = 0.93$; hence $c_{\rm R} = 2.4 \! \times \! 10^5 \, {\rm cm \, s}^{-1}.$ Using (\ref{R modes}),
\begin{equation}
{\bm \nabla} \cdot {\bf f}_{\rm R} = {K^{3 \over 2} (\varphi^2 -1) \over \sqrt{{\cal K}A} } \, e^{i {\bf K} \cdot {\bf r}} \, e^{- \varphi K z}
\label{R divergence}
\end{equation}
and $U_{\rm R} = (2 \pi c_{\rm R}^4 K / {\cal K} \, v_{\rm l}^4 V_{\rm el}) \, I_{\rm R}(Kd),$ where
$$
I_{\rm R}(Z) \! \equiv \! \int_0^Z \! \! dx \, dx' \, {\sf K}_0\big( \sqrt{(x \! - \! x')^2 + a^2Z^2/d^2 } \big) e^{-\varphi (x+x')}.
$$

The final summations in (\ref{continuous weighted DOS}) are carried out with the aid of the identity $\lim_{A \rightarrow \infty} \sum_{\bf K} \delta(\omega - cK) = \omega A /2 \pi c^2$ and elsewhere replacing $K$ with $\omega/c$. Then we obtain
\begin{eqnarray}
F(\omega) &=&  \frac{\omega^2}{ v_{\rm l}^4 d} \bigg\lbrace  {c_{\rm R} \over {\cal K}} \, I_{\rm R}({\textstyle{\omega d \over c_{\rm R}}}) 
+ \int_{v_{\rm t}}^{v_{\rm l}} \! \! dc \ {|{\cal C}|^2 \over {2 \pi \beta}} \, I_{0}({\textstyle{\omega d \over c}}, c) \nonumber \\
&+&   \int_{v_{\rm l}}^\infty \! \! \! dc \ {1 \over 2 \pi \alpha} \big[ I_{+}({\textstyle{\omega d \over c}}, c) +  I_{-}({\textstyle{\omega d \over c}}, c) \big] \bigg\rbrace.
\label{collected F}
\end{eqnarray}
This expression, combined with (\ref{thermal power formula}), is our principal result. Evaluation of (\ref{collected F}) can be further simplified by the use of the powerful identities
\begin{eqnarray}
I_\pm(Z,c) &=& {\rm Re} \big[ \big( 2Z - {\textstyle{i \zeta_\pm \over \alpha}} \big) f(Z,\alpha) + {\textstyle{i \zeta_\pm \over \alpha}} e^{2 i \alpha Z}  f^*(Z,\alpha) \nonumber \\
&+& 2 i \big( {\textstyle{\partial f(Z,s) \over \partial s}} \big)_{s=\alpha} \big], \\
I_0(Z,c) &=& {\textstyle{1 \over \gamma}} \, f(Z, i \gamma) -  {\textstyle{e^{-2 \gamma Z} \over \gamma}}  \, f(Z,- i \gamma) , \\
I_{\rm R}(Z,c) &=& {\textstyle{1 \over \varphi}} \, f(Z, i \varphi) - {\textstyle{ e^{-2 \varphi Z} \over \varphi}} \, f(Z,- i \varphi) ,
\end{eqnarray}
where $f(Z,s) \equiv \int_0^Z \! dx \, K_0\big(\sqrt{x^2 + a^2 Z^2/d^2}\big) \, e^{i s x},$ thereby reducing the $I_m$ to a single one-dimensional integral $f.$

The $I_m$ have distinct large- and small-$Z$ character, crossing over near $Z=1$.  Because of the integration over $c$ in (\ref{collected F}), $F$ and $P$ accordingly exhibit a broad crossover behavior. However, once $\omega d  < c_{\rm R}$, all branches will have assumed their low-frequency forms. We define a crossover temperature
\begin{equation} 
T^\star \equiv \hbar c_{\rm R} / k_{\rm B} d
\end{equation}
dividing regimes determined by the small and large $\omega d / c_{\rm R}$ behavior of $F$. In the large $\omega d / c_{\rm R}$ limit the $m \! = \! \pm$ modes in (\ref{collected F}) can be shown to be dominant, and $\lim_{\omega d \rightarrow \infty} \int_{v_{\rm l}}^\infty \! \! dc \ \frac{1}{\alpha}  I_{\pm}({\textstyle{\omega d \over c}}, c) = \pi \omega d.$ Therefore, we obtain $F(\omega) \rightarrow F_{\rm bulk}(\omega) \equiv \omega^3 / v_{\rm l}^4,$ independent of $d$, leading to a high-temperature behavior $P \rightarrow \Sigma \, V_{\rm el} \big[ \Phi(\omega_{\rm D}/T_{\rm el}) \, T_{\rm el}^5 - \Phi(\omega_{\rm D}/T_{\rm ph}) \, T_{\rm ph}^5 \big],$ where $\Sigma$ is the coefficient (\ref{Wellstood coefficient}), and where $\Phi(y) \equiv [4! \, \zeta(5)]^{-1} \! \int_0^y dx \ x^4 /(e^x - 1).$ $\Phi(10)$ is about 0.97, and $\Phi(y)$ rapidly approaches 1 beyond that. Thus, at temperatures above $T^\star$ but sufficiently smaller than the Debye temperature, the $\Phi$ factors are equal to unity, and we recover the bulk result (\ref{Wellstood formula}).

The low temperature asymptotic analysis is somewhat complicated and will be presented elsewhere. Briefly, using the small $Z$ expansion
\begin{eqnarray}
f(Z,s) &\rightarrow& - Z \ln Z + \big(1 + \ln 2 + \psi(1) \big)Z - \frac{i s}{2} Z^2 \ln Z \nonumber \\
&+& \frac{i s}{2}\big({\textstyle{1 \over 2}} + \ln 2 + \psi(1) \big) Z^2 + O(Z^3 \ln Z),
\end{eqnarray}
where $\psi$ is the Euler polygamma function, we find $F(\omega) \rightarrow F_{\rm bulk}(\omega) \! \times \! [ - \lambda  \, ({\textstyle{\omega \, d \over c_{\rm R}}})  \ln  ({\textstyle{\omega \, d \over c_{\rm R}}}) + O({\textstyle{\omega \, d \over c_{\rm R}}})]$ in the small $\omega d / c_{\rm R}$ limit. Here
\begin{eqnarray*}
\lambda &\equiv& \frac{1}{\cal K} +  \int_{v_{\rm t}}^{v_{\rm l}} \! \! \! dc \, \frac{c_{\rm R} |{\cal C}|^2}{2 \pi c^2 \beta} + \int_{v_{\rm l}}^\infty  \! \! \! dc \, 
\frac{c_{\rm R}[2 \! - \! {\rm Re}(\xi_{+} \! + \! \xi_{-})]}{2 \pi c^2 \alpha}
\end{eqnarray*}
is a constant determined by $v_{\rm l}$, $v_{\rm t},$ and $c_{\rm R}.$ Each $T^5$ function in (\ref{Wellstood formula}) therefore crosses over at low temperature as $T^5 \rightarrow - \Lambda  \big({\textstyle{T^6 \over T^\star}}\big) \ln \big({\textstyle{T \over T^\star}}\big),$ with $\Lambda = {\lambda \, \pi^6 / 189 \, \zeta(5)}.$ For a Cu film, $\lambda \approx 0.815$ and $\Lambda \approx 3.998$. There are also mixed-temperature regimes possible, where only one of the two terms in (\ref{Wellstood formula}) has crossed over.

The most striking consequence of the crossover is that the temperature exponent increases. In Fig.~\ref{exponent figure} we fit $P$ (with either $T_{\rm el}$ or $T_{\rm ph}$ zero) to a power-law $T^x$ with a temperature dependent exponent $x$, and plot the exponent for $10 \, {\rm nm}$ ($T^\star = 1.84 \, {\rm K}$) and  $100 \, {\rm nm}$ ($T^\star = 184 \, {\rm mK}$) Cu films. $x(T)$ is nonmonatonic, displaying a pronounced maximum near $T^\star$, and drifts upward as $T \rightarrow 0.$ Such behavior has not (to the best of our knowledge) been observed, even though many experiments \cite{WellstoodPRB94,RoukesPRL85,YungAPL02,DiTusaPRL92} have achieved $T \ll T^\star \! .$ The physical origin of the crossover is that, at low temperature, the stress-free condition at the metal surface penetrates into the film, reducing the strain and hence electron-phonon coupling there. The characteristic distance over which the boundary condition has an effect is of the order of a bulk wavelength. When $T \gg T^\star$, only a thin outer surface layer of the film has a significantly diminished strain, and bulk behavior is expected. However, when $T \ll T^\star$ the entire metal film experiences a reduced strain. 

The experiments of Refs.~\cite{RoukesPRL85} and \cite{YungAPL02}, both using Cu films, observe an approximate $T^5$ dependence even well below $T^\star \! .$ It is therefore interesting to compare the observed {\it prefactors} with the coefficient $\Sigma,$ evaluated for Cu. Using a free-electron gas approximation \cite{freenote} and measured elastic properties \cite{elasticnote}, we obtain $ 5.97 \! \times \! 10^7 \, {\rm W \, m^{-3} \, K^{-5}},$ which is at least an order of magnitude smaller than observed, consistent with our assertion that there is some unidentified mechanism {\it enhancing} the  thermal coupling. Noble metals are far from free-electron systems because of their complex Fermi surfaces. We attempt to address this shortcoming by regarding the ``Fermi surface'' quantities  $N_{\rm el}(\epsilon_{\rm F})$ and $v_{\rm F}$ as independently adjustable parameters, to be obtained empirically from heat capacity and cyclotron resonance data. Carrying out this analysis, the details of which will be presented elsewhere,  leads to the modified prefactor $\Sigma = 1.14 \! \times \! 10^8 \, {\rm W \, m^{-3} \, K^{-5}},$ which is still considerably smaller than measured.

Although not included in the model considered here, disorder in a bulk metal film is expected to produce a crossover from the $T^5$ dependence to a $T^6$ scaling when the phonon elastic mean free path $\ell$ becomes smaller than the thermal wavelength \cite{KeckJLTP76,RammerPRB86}, a behavior which has not been reported experimentally until very recently \cite{MaasiltaPre04}. Thus, the  crossover predicted here should not be appreciable affected by disorder unless $\ell < d$. Although thin films are known to scatter phonons strongly, measured values of $\ell$ are still much larger than $d$ in the temperature regime of interest here \cite{KlitsnerPRB87}.

In conclusion, we argue that a wide variety of experiments contradict the predictions of an essentially exact application of the standard model of electron-phonon thermal coupling in metals to a supported-film geometry, suggesting a widespread breakdown of that model. ANC was supported by the NASA Office of Space Science under grant NAG5-8669, and by the Army Research Office under DAAD-19-99-1-0226. MRG was supported by the National Science Foundation under grants DMR-0093217 and CMS-040403.

\bibliography{/Users/mgeller/Papers/bibliographies/MRGpre,/Users/mgeller/Papers/bibliographies/MRGbooks,/Users/mgeller/Papers/bibliographies/MRGgroup,/Users/mgeller/Papers/bibliographies/MRGphonons,/Users/mgeller/Papers/bibliographies/MRGcm,/Users/mgeller/Papers/bibliographies/MRGnano,hotnotes}

\begin{thebibliography}{20}
\expandafter\ifx\csname natexlab\endcsname\relax\def\natexlab#1{#1}\fi
\expandafter\ifx\csname bibnamefont\endcsname\relax
  \def\bibnamefont#1{#1}\fi
\expandafter\ifx\csname bibfnamefont\endcsname\relax
  \def\bibfnamefont#1{#1}\fi
\expandafter\ifx\csname citenamefont\endcsname\relax
  \def\citenamefont#1{#1}\fi
\expandafter\ifx\csname url\endcsname\relax
  \def\url#1{\texttt{#1}}\fi
\expandafter\ifx\csname urlprefix\endcsname\relax\def\urlprefix{URL }\fi
\providecommand{\bibinfo}[2]{#2}
\providecommand{\eprint}[2][]{\url{#2}}

\bibitem[{\citenamefont{Gantmakher}(1974)}]{GantmakherRPP74}
\bibinfo{author}{\bibfnamefont{V.~F.} \bibnamefont{Gantmakher}},
  \bibinfo{journal}{Rep. Prog. Phys.} \textbf{\bibinfo{volume}{37}},
  \bibinfo{pages}{317} (\bibinfo{year}{1974}).

\bibitem[{\citenamefont{Wellstood et~al.}(1994)\citenamefont{Wellstood, Urbina,
  and Clarke}}]{WellstoodPRB94}
\bibinfo{author}{\bibfnamefont{F.~C.} \bibnamefont{Wellstood}},
  \bibinfo{author}{\bibfnamefont{C.}~\bibnamefont{Urbina}}, \bibnamefont{and}
  \bibinfo{author}{\bibfnamefont{J.}~\bibnamefont{Clarke}},
  \bibinfo{journal}{Phys. Rev. B} \textbf{\bibinfo{volume}{49}},
  \bibinfo{pages}{5942} (\bibinfo{year}{1994}).

\bibitem[{\citenamefont{Little}(1959)}]{Little59}
\bibinfo{author}{\bibfnamefont{W.~A.} \bibnamefont{Little}},
  \bibinfo{journal}{Can. J. Phys.} \textbf{\bibinfo{volume}{37}},
  \bibinfo{pages}{334} (\bibinfo{year}{1959}).

\bibitem[{\citenamefont{Roukes et~al.}(1985)\citenamefont{Roukes, Freeman,
  Germain, Richardson, and Ketchen}}]{RoukesPRL85}
\bibinfo{author}{\bibfnamefont{M.~L.} \bibnamefont{Roukes}},
  \bibinfo{author}{\bibfnamefont{M.~R.} \bibnamefont{Freeman}},
  \bibinfo{author}{\bibfnamefont{R.~S.} \bibnamefont{Germain}},
  \bibinfo{author}{\bibfnamefont{R.~C.} \bibnamefont{Richardson}},
  \bibnamefont{and} \bibinfo{author}{\bibfnamefont{M.~B.}
  \bibnamefont{Ketchen}}, \bibinfo{journal}{Phys. Rev. Lett.}
  \textbf{\bibinfo{volume}{55}}, \bibinfo{pages}{422} (\bibinfo{year}{1985}).

\bibitem[{\citenamefont{Yung et~al.}(2002)\citenamefont{Yung, Schmidt, and
  Cleland}}]{YungAPL02}
\bibinfo{author}{\bibfnamefont{C.~S.} \bibnamefont{Yung}},
  \bibinfo{author}{\bibfnamefont{D.~R.} \bibnamefont{Schmidt}},
  \bibnamefont{and} \bibinfo{author}{\bibfnamefont{A.~N.}
  \bibnamefont{Cleland}}, \bibinfo{journal}{Appl. Phys. Lett.}
  \textbf{\bibinfo{volume}{81}}, \bibinfo{pages}{31} (\bibinfo{year}{2002}).

\bibitem[{\citenamefont{Liu and Giordano}(1991)}]{LiuPRB91}
\bibinfo{author}{\bibfnamefont{J.}~\bibnamefont{Liu}} \bibnamefont{and}
  \bibinfo{author}{\bibfnamefont{N.}~\bibnamefont{Giordano}},
  \bibinfo{journal}{Phys. Rev. B} \textbf{\bibinfo{volume}{43}},
  \bibinfo{pages}{3928} (\bibinfo{year}{1991}).

\bibitem[{\citenamefont{DiTusa et~al.}(1992)\citenamefont{DiTusa, Lin, Park,
  Isaacson, and Parpia}}]{DiTusaPRL92}
\bibinfo{author}{\bibfnamefont{J.~F.} \bibnamefont{DiTusa}},
  \bibinfo{author}{\bibfnamefont{K.}~\bibnamefont{Lin}},
  \bibinfo{author}{\bibfnamefont{M.}~\bibnamefont{Park}},
  \bibinfo{author}{\bibfnamefont{M.~S.} \bibnamefont{Isaacson}},
  \bibnamefont{and} \bibinfo{author}{\bibfnamefont{J.~M.}
  \bibnamefont{Parpia}}, \bibinfo{journal}{Phys. Rev. Lett.}
  \textbf{\bibinfo{volume}{68}}, \bibinfo{pages}{1156} (\bibinfo{year}{1992}).

\bibitem[{\citenamefont{Roukes}(1999)}]{Roukes99}
\bibinfo{author}{\bibfnamefont{M.~L.} \bibnamefont{Roukes}},
  \bibinfo{journal}{Physica B} \textbf{\bibinfo{volume}{263}},
  \bibinfo{pages}{1} (\bibinfo{year}{1999}).

\bibitem[{\citenamefont{Schwab et~al.}(2000)\citenamefont{Schwab, Henriksen,
  Worlock, and Roukes}}]{SchwabNat00}
\bibinfo{author}{\bibfnamefont{K.}~\bibnamefont{Schwab}},
  \bibinfo{author}{\bibfnamefont{E.~A.} \bibnamefont{Henriksen}},
  \bibinfo{author}{\bibfnamefont{J.~M.} \bibnamefont{Worlock}},
  \bibnamefont{and} \bibinfo{author}{\bibfnamefont{M.~L.}
  \bibnamefont{Roukes}}, \bibinfo{journal}{Nature (London)}
  \textbf{\bibinfo{volume}{404}}, \bibinfo{pages}{974} (\bibinfo{year}{2000}).

\bibitem[{\citenamefont{Schmidt et~al.}(2004)\citenamefont{Schmidt, Schoelkopf,
  and Cleland}}]{SchmidtPRL04}
\bibinfo{author}{\bibfnamefont{D.~R.} \bibnamefont{Schmidt}},
  \bibinfo{author}{\bibfnamefont{R.~J.} \bibnamefont{Schoelkopf}},
  \bibnamefont{and} \bibinfo{author}{\bibfnamefont{A.~N.}
  \bibnamefont{Cleland}}, \bibinfo{journal}{Phys. Rev. Lett.}
  \textbf{\bibinfo{volume}{93}}, \bibinfo{pages}{45901} (\bibinfo{year}{2004}).

\bibitem[{den()}]{densitynote}
\bibinfo{note}{Recall that, in elasticity theory, the mass-density fluctuation
  $\delta \rho$ is given by $- \rho {\bm \nabla} \cdot {\bf u}$.}

\bibitem[{\citenamefont{Allen}(1987)}]{AllenPRL87}
\bibinfo{author}{\bibfnamefont{P.~B.} \bibnamefont{Allen}},
  \bibinfo{journal}{Phys. Rev. Lett.} \textbf{\bibinfo{volume}{59}},
  \bibinfo{pages}{1460} (\bibinfo{year}{1987}).

\bibitem[{\citenamefont{Ezawa}(1971)}]{Ezawa71}
\bibinfo{author}{\bibfnamefont{H.}~\bibnamefont{Ezawa}}, \bibinfo{journal}{Ann.
  Phys. (N.Y.)} \textbf{\bibinfo{volume}{67}}, \bibinfo{pages}{438}
  (\bibinfo{year}{1971}).

\bibitem[{fre()}]{freenote}
\bibinfo{note}{In the free-electron gas approximation, the Fermi energy of Cu
  is $7.03 \, {\rm eV}$, the DOS at $\epsilon_{\rm F} $ is $1.13 \times 10^{34}
  \, {\rm erg^{-1} \, cm^{-3}}$, and $v_{\rm F}$ is $1.57 \times 10^{8} \, {\rm
  cm \, s^{-1}}$}.

\bibitem[{ela()}]{elasticnote}
\bibinfo{note}{Cu has the low-temperature elastic constants $c_{11} = 1.76
  \times 10^{12} \, {\rm dyne \, cm^{-2}} \! ,$ $c_{12} = 1.25 \times 10^{12}
  \, {\rm dyne \, cm^{-2}} \! ,$ and $c_{44} = 8.18 \times 10^{11} \, {\rm dyne
  \, cm^{-2}}$ \cite{OvertonPR55}, and its mass density is $\rho = 8.94 \, {\rm
  g \, cm^{-3}}$. To approximate a cubic material as an isotropic elastic
  continuum we define effective elastic constants $C_{11} \equiv c_{11} - 2
  \Delta$ and $C_{44} \equiv c_{44} + \Delta$, where $\Delta \equiv (c_{11} -
  c_{12} - 2 c_{44})/5$, and we let $v_{\rm l} = (C_{11}/\rho)^{1 \over 2}$ and
  $v_{\rm t} = (C_{44}/\rho)^{1 \over 2}$. This leads to $v_{\rm l} = 4.97
  \times 10^{5} \, {\rm cm \, s^{-1}}$}.

\bibitem[{\citenamefont{Keck and Schmid}(1976)}]{KeckJLTP76}
\bibinfo{author}{\bibfnamefont{B.}~\bibnamefont{Keck}} \bibnamefont{and}
  \bibinfo{author}{\bibfnamefont{A.}~\bibnamefont{Schmid}},
  \bibinfo{journal}{J. Low. Temp. Phys.} \textbf{\bibinfo{volume}{24}},
  \bibinfo{pages}{611} (\bibinfo{year}{1976}).

\bibitem[{\citenamefont{Rammer and Schmid}(1986)}]{RammerPRB86}
\bibinfo{author}{\bibfnamefont{J.}~\bibnamefont{Rammer}} \bibnamefont{and}
  \bibinfo{author}{\bibfnamefont{A.}~\bibnamefont{Schmid}},
  \bibinfo{journal}{Phys. Rev. B} \textbf{\bibinfo{volume}{34}},
  \bibinfo{pages}{1352} (\bibinfo{year}{1986}).

\bibitem[{\citenamefont{Maasilta et~al.}()\citenamefont{Maasilta, Karvonen,
  Kivioja, and Taskinen}}]{MaasiltaPre04}
\bibinfo{author}{\bibfnamefont{I.~J.} \bibnamefont{Maasilta}},
  \bibinfo{author}{\bibfnamefont{J.~T.} \bibnamefont{Karvonen}},
  \bibinfo{author}{\bibfnamefont{J.~M.} \bibnamefont{Kivioja}},
  \bibnamefont{and} \bibinfo{author}{\bibfnamefont{L.~J.}
  \bibnamefont{Taskinen}}, \bibinfo{note}{e-print cond-mat/0311031}.

\bibitem[{\citenamefont{Klitsner and Pohl}(1987)}]{KlitsnerPRB87}
\bibinfo{author}{\bibfnamefont{T.}~\bibnamefont{Klitsner}} \bibnamefont{and}
  \bibinfo{author}{\bibfnamefont{R.~O.} \bibnamefont{Pohl}},
  \bibinfo{journal}{Phys. Rev. B} \textbf{\bibinfo{volume}{36}},
  \bibinfo{pages}{6551} (\bibinfo{year}{1987}).

\bibitem[{\citenamefont{Overton and Gaffney}(1955)}]{OvertonPR55}
\bibinfo{author}{\bibfnamefont{W.~C.} \bibnamefont{Overton}} \bibnamefont{and}
  \bibinfo{author}{\bibfnamefont{J.}~\bibnamefont{Gaffney}},
  \bibinfo{journal}{Phys. Rev.} \textbf{\bibinfo{volume}{98}},
  \bibinfo{pages}{969} (\bibinfo{year}{1955}).

\end{thebibliography}

\end{document}